\documentclass[mathleft]{an}
\usepackage{graphicx}
\usepackage{times}
\usepackage[]{caption}

\usepackage{natbib}
\bibpunct{(}{)}{;}{a}{}{,}
\begin{document}
\Pagespan{1}{}

\Yearpublication{2013}%
\Yearsubmission{2013}%
\Month{1}%
\Volume{1}%
\Issue{1}%

\title{STILT: System Design $\&$ Performance}

\author{N.R.Mawson\inst{1}\fnmsep\thanks{Corresponding author:
  \email{nrm@astro.livjm.ac.uk}\newline}
\and  I.A.Steele\inst{1}
\and R.J.Smith\inst{1}
}
\titlerunning{STILT pipeline}
\authorrunning{N.R. Mawson \& I.A. Steele \& R.J. Smith}
\institute{
Astrophysics Research Institute, Liverpool John Moores University, Egerton Wharf, Birkenhead, CH41 1LD}

\received{4 Mar 2013}
\accepted{1 May 2013}
\publonline{later}


\abstract{%
  The Small Telescopes Installed at the Liverpool Telescope (STILT) have been in operation since March 2009, collecting wide field data from their position, mounted to the Liverpool Telescope.  The two instruments; SkycamT and SkycamZ have been used to create a variability search of the skies visible at La Palma with the limits of 12$^{th}$ and 18$^{th}$ R band magnitude with fields of view of 21x21$^o$ and 1x1$^o$.  We provide here a description of the hardware and software setup and the performance of the system to date.}

\maketitle

\section{Introduction}
In the last 15 years there has been an explosion in the number of small aperture observing systems consisting of either small telescopes, or simply arrays of widely available camera lens' attached to large format CCDs.  These systems are generally very cost effective and can generate large volumes of data and results, for a relatively small startup cost.  The main science areas they provide for are in the time domain, which up until the advent of these systems, was a relatively unknown parameter space.  They generally provide wide observing fields, which are invaluable for sky surveys, as they have the ability to cover large areas of the sky quickly.  Systems such as OGLE \citep{OGLE}, MACHO \citep{MACHO}, ASAS \citep{ASAS}, ROTSE \citep{ROTSE} and PI of the sky \citep{PI} have shown just how valuable wide field surveys and sky monitoring can be.  In this paper we describe the setup, calibration, data reduction and performance of a pair of wide field imaging systems (STILT), installed at the Liverpool Telescope on La Palma.  Its objective is to provide complimentary followup of LT observations and to carry out a serendipitous search for variable stars.

\section{STILT System Description}
The Liverpool Telescope (LT) is a fully robotic 2m telescope based at the Observatorio del Roque de los Muchachos on La Palma \citep{LT}.  Since March 2009 the Small Telescopes Installed at the LT (STILT) have been amassing nightly unfiltered photometric data.

STILT consists of two separate imaging cameras each individually identified as a Skycam. SkycamZ consists of an Orion Optics AG8 astrograph, which has a 200mm primary mirror, 760mm focal length, and focal ratio f/3.8. It has been modified to support an extra length tube baffle to reduce stray light and a strengthened focussing mechanism to improve image stability.  These additions were added in September and November 2010.  

SkycamT consists of a Nikon 35mm lens set at f/4.  Both optical systems are paired with an Andor ikon-M DU934N-BV CCD which provide a 1024x1024 pixel, back-illuminated and anti-reflection coated detector with a pixel size of 13$\mu$m. This gives fields of view of 1x1 degrees and 21x21 degrees respectively.  The cameras are attached to the secondary mirror support structure of the LT, which means they parallel point with the main telescope, as shown in Figure 1.  The CCDs remain unfiltered to maximise throughput.  Their quantum efficiency curve is shown in Figure 2.  

\begin{figure}[ht]
\centering
\includegraphics[width=0.45\textwidth]{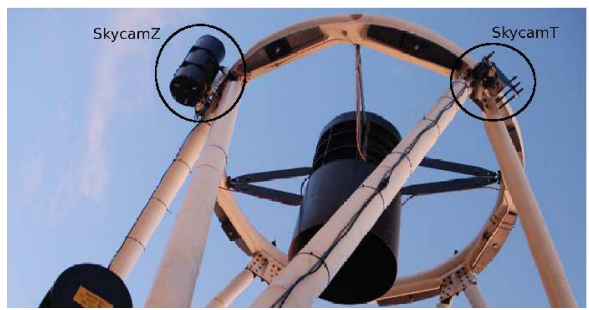}
\caption{The Liverpool Telescope with SkycamZ and SkycamT attached to the secondary mirror support.}
\end{figure}

\begin{center}
\begin{table*}[!ht]
\caption{Table of Skycam specifications}
\hfill{}
\begin{tabular}{|ll|}
\hline
\hline
\multicolumn{1}{|c}{Characteristic} & \multicolumn{1}{c|}{Value} \\
\cline{1-2}
\hline
\multicolumn{2}{|c|}{SkycamZ} \\
\hline
&\\
Telescope.............................. & Orion Optics AG8 astrograph 200mm \\
Field of View....................... & 1$^o$ x 1$^o$\\
Pixel Scale........................... & 3.5'' pixel$^{-1}$\\
Limits.................................. & 18th USNO-B R-band magnitude\\
Gain.................................... & 5.3 e$^-$ per ADU\\
Read noise........................... & 1.0$\pm$0.2 ADU\\
Read speed.......................... & 35sec\\
&\\
\hline
\multicolumn{2}{|c|}{SkycamT} \\
\hline
&\\
Optics.................................. & Nikon 35mm F/2 (set to f/4) \\
Field of View ...................... & 21$^o$ x 21$^o$\\
Pixel Scale........................... & 75.6'' pixel$^{-1}$\\
Limits.................................. & 12th USNO-B R-band magnitude\\
Gain.................................... & 5.2 e$^-$ per ADU\\
Read noise........................... & 1.1$\pm$0.2 ADU\\
Read speed.......................... & 35sec\\
&\\
\hline
\multicolumn{2}{|c|}{Andor CCDs} \\
\hline
&\\
Model.................................. & Andor ikon-M DU934N-BV\\
Active Pixels....................... & 1024x1024\\
Pixel Size............................ & 13$\mu$m\\
Image area.......................... & 13.3 x 13.3mm\\
\hline
\end{tabular}
\hfill{}
\end{table*}
\end{center}

\begin{figure}[ht]
\centering
\includegraphics[width=0.5\textwidth]{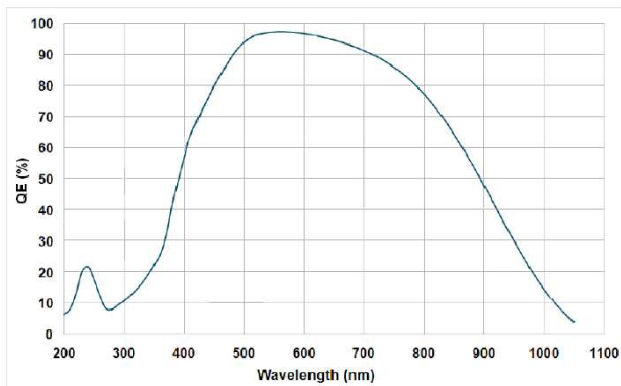}
\caption{Quantum efficiency of the Andor cameras, taken from the Andor iKon-M 934 specifications booklet.  As the cameras are unfiltered, this is the response of the CCDs.}
\end{figure}

Each Skycam has its own computer control system, comprising of an Asus EeeBox mini desktop PC, which provides the CCD controls and automated scripts. A ``cron'' script is set to carry out 10s integrations every minute whenever the telescope dome is open.  This in turn runs a C script which controls the CCD and contains the command to take an exposure.  The resultant detection limits of the cameras are 18th and 12th magnitude in the R band for SkycamZ and SkycamT respectively.  The setup of these cameras will be discussed further in this paper.  
  
A data reduction pipeline is in place to carry out data reduction tasks.  This involves the standard image reduction methods but modified for the large field size of images. This is discussed further in section III.
A data analysis pipeline has been used to process the archive of images for both cameras and create a database of the resulting data. This and the associated database is discussed in section IV.
The performance of the system in terms of photometry and astrometry is discussed in section V.

A followup paper (\citealt{stilt2}, in prep), will cover the exploitation of the finalised database and the results inferred from it.

\subsection{Data Summary}

The system of 10s observations, every minute for a night's duration provides on average 750 frames per night per camera, all of which are stored centrally in the LT archive.  The data used to compile the STILT database was taken from the period of commissioning; March 2009 for SkycamT and July 2009 for SkycamZ, to the end of March 2012. The details of this dataset are held in Table 2.

Spatially, as the STILT system has no control over its pointing, the system covers a wide area of the sky to varying levels of repetition.  SkycamT has covered most of the observable sky at La Palma, while SkycamZ provides ``pin head'' repeat observations across the sky but by no means provides a complete sky coverage to its limiting magnitude. 
 
\begin{center}
\begin{figure*}[!ht]
\includegraphics[width=\textwidth]{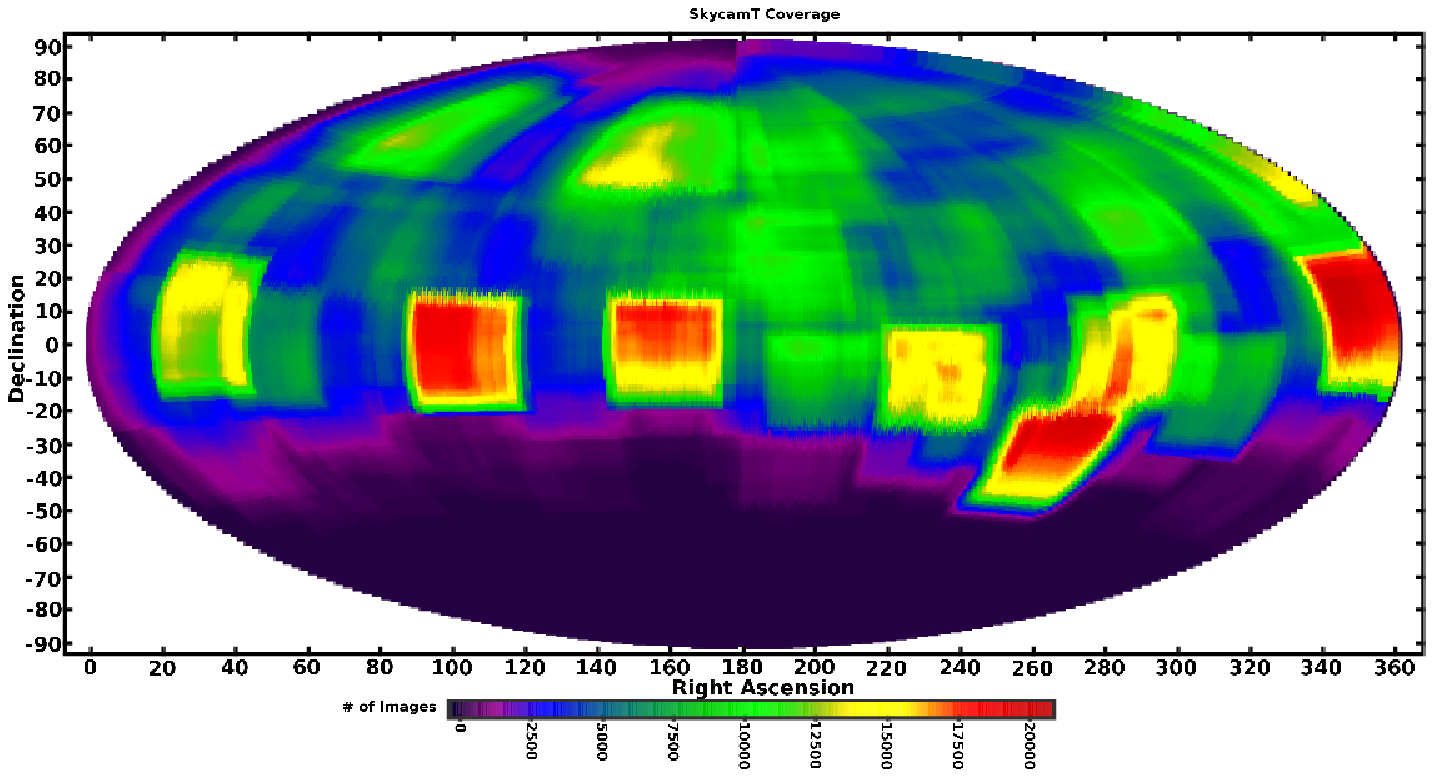}
\caption{Hammer projection of the number of fields observed by SkycamT during its first 3 years}
\includegraphics[width=\textwidth]{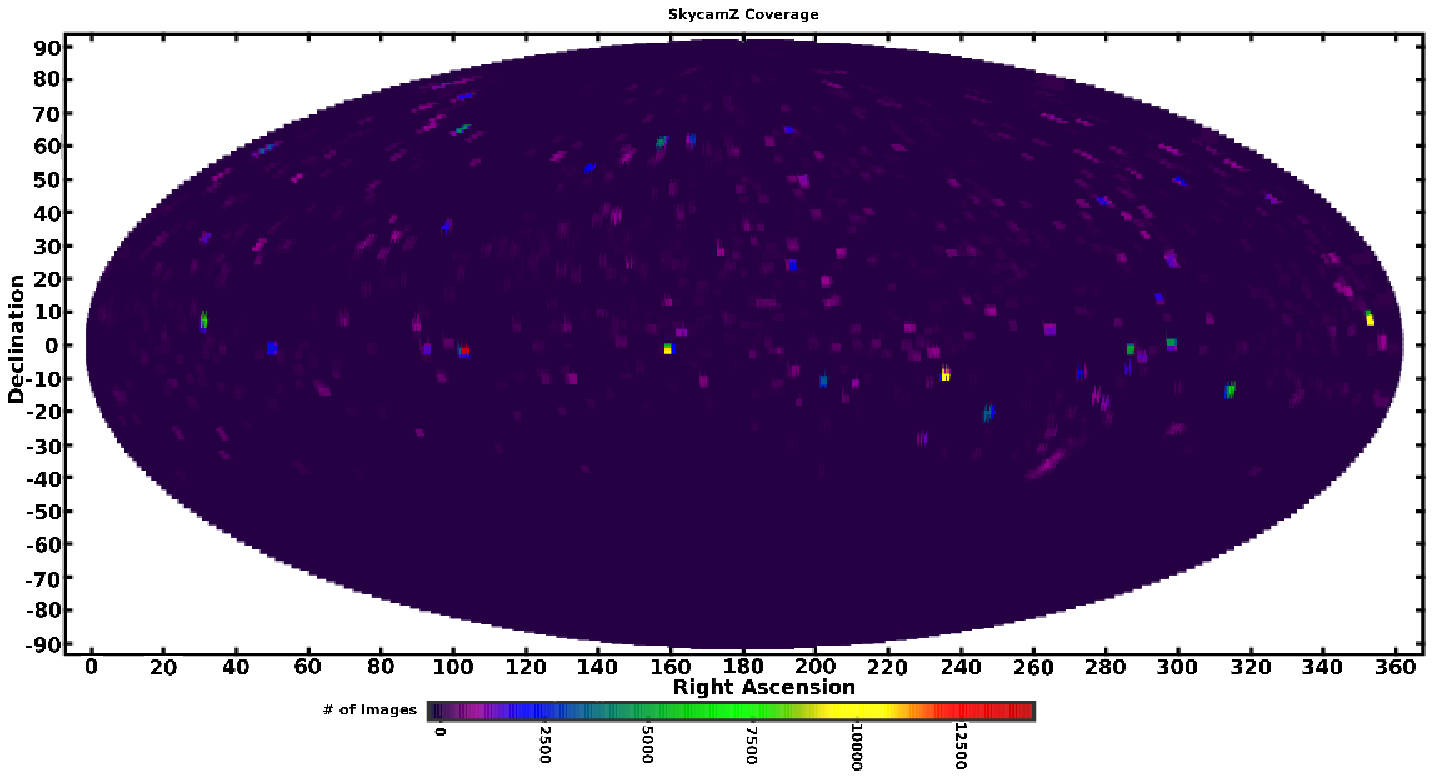}
\caption{Hammer projection of the number of fields observed by SkycamZ during its first 3 years}
\end{figure*}
\end{center}

The coverage of the systems is displayed in Figures 3 and 4, the counts per object for each camera is shown in Figure 5 and the number of observations per field in Figures 6 and 7.

\section{Image reduction}
The image reduction framework carries out the necessary data reduction and world coordinate system fitting to the science images.  It uses standard methodology for photometric data except for some modifications to take into account the large fields of view of the cameras.

\begin{center}
\begin{table}
\caption{Summary of data}
\begin{tabular}{lcc}
\hline
 & SkycamZ & SkycamT\\
\hline\\
Images & 272,470 & 315,277\\
Unique Objects & 6,290,935 & 21,453,608\\
Data points & 332,735,320 & 904,033,139\\
Mean LC points & 53 & 42 \\
Median LC points & 1 & 7 \\
Maximum LC points & 13802 & 17370 \\
Image disk space & \multicolumn{2}{c}{1.0 Tb} \\
Database disk space & 66.4 Gb & 213 Gb \\
Start date & 29/06/2009 & 05/03/2009 \\
End date & 31/03/2012 & 31/03/2012 \\
Minimum Declination & -36.0577 & -51.1956\\
Maximum Declination & 87.8928 & 90\\
\hline
\end{tabular}
\end{table}
\end{center}

\begin{figure}[!ht]
\centering
\includegraphics[width=0.5\textwidth]{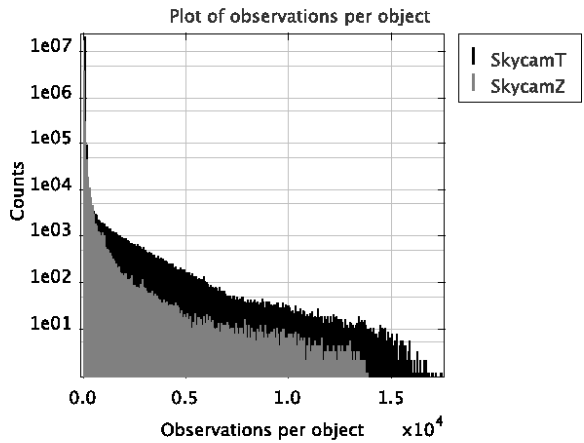}
\caption{Histogram of the number of observations for each unique object in both the SkycamT and SkycamZ database}
\end{figure}

\begin{center}
\begin{figure}[!ht]
\includegraphics[width=0.485\textwidth]{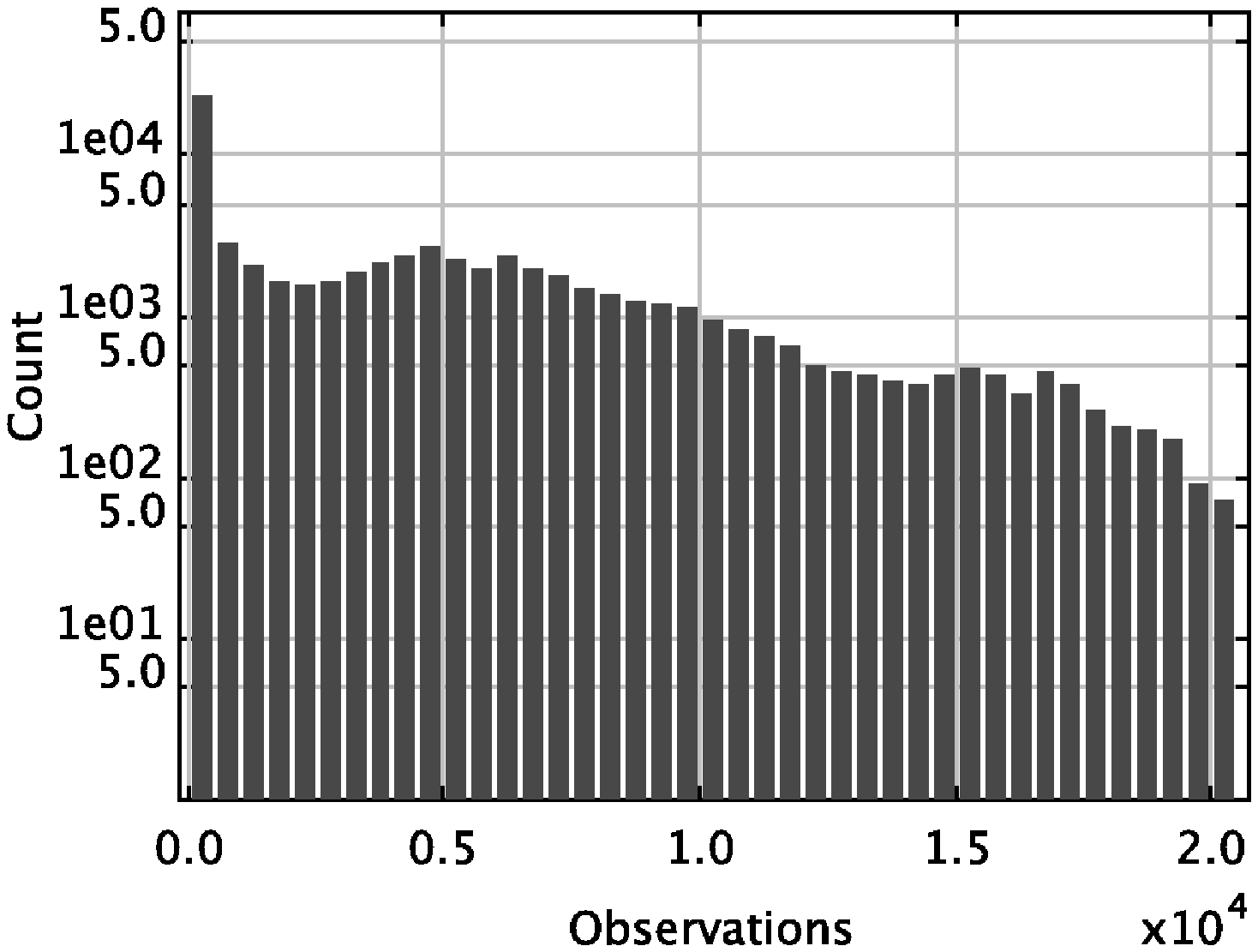}
\caption{Histogram of the number of observations per field in SkycamT}
\includegraphics[width=0.485\textwidth]{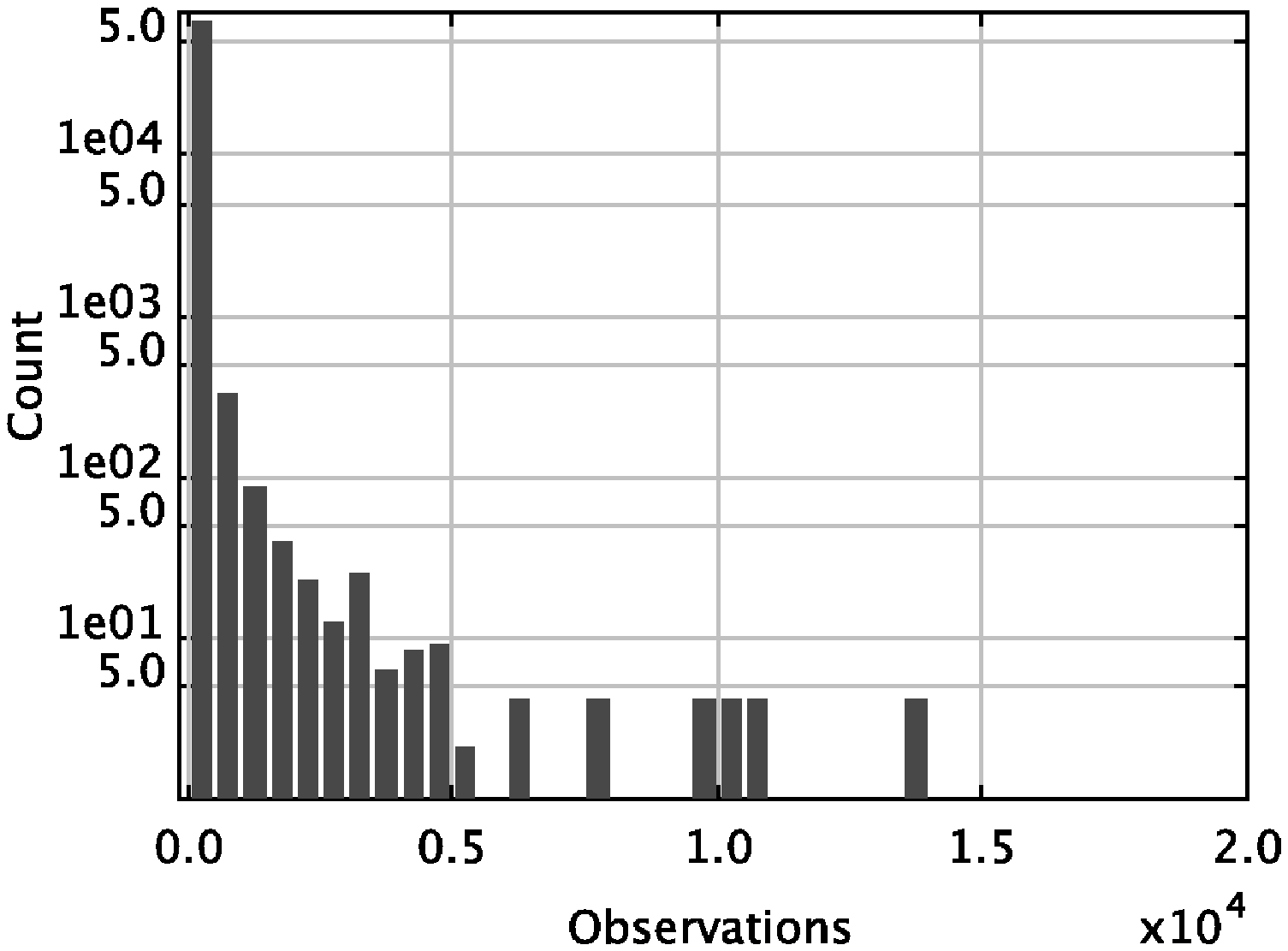}
\caption{Histogram of the number of observations per field in SkycamZ}
\end{figure}
\end{center}

Dark current and bias signal are removed in one step as all the frames produced by the Skycams are of the same exposure time.  The effects are removed by producing a master dark frame by stacking between 30-210 dark frames, which are updated on a weekly basis.

Flat fielding is normally done by taking twilight images which give a good approximation to a uniformly illuminated CCD, however due to the large area of sky covered by each image taken by a Skycam variation in the sky becomes an important effect.  Due to this a stacked flat is used for the Skycams.  This consists of a stack of between 500-1000 twilight flats taken over periods of two months. Each individual flat is divided through by a 'plane surface' to remove the gradients caused by the large area of sky being imaged.  This stacked flat is then used to remove the variations across the detector.  The flats are updated on a monthly basis.

\subsection{World coordinate system}
Astrometry.net \citep{Astrometry} is a system that produces all the necessary data for an astrometric fit; pointing, scale, and orientation.  It requires no ``first guess'' and works from information in the image pixels alone.  \citet{Astrometry} claim a $>$ 99$\%$ success rate for contemporary near-ultraviolet and visual imaging survey data, with no false positives for fields with a well matched set of reference quads.  The Skycams exist on an independent network, for security reasons, so are not linked to any part of the LTs computer system.  Therefore no pointing information is transferred to the Skycam images.  Using Astrometry.net means we do not need a direct astrometric link to the main telescope, this makes the Astrometry.net software ideal for the fitting of a WCS to the Skycam data.

The applied WCS is a tangent plane projection (TAN) for SkycamZ.  SkycamT fits also include extra 3$^{rd}$ order Simple Imaging Polynomial (SIP) terms \citep{sip} to account for the distortion induced by the wide field of view.  These polynomials provide a means to represent the non-linear distortions caused by the large field of view inherent in SkycamT.

We find a similar success rate with both Skycams, with SkycamT showing a 99.87$\%$ success rate for a subset of three months data taken at the start of 2012.  This result is true for fields which are not taken during a slew, are not $>$10$\%$ saturated and $<$50$\%$ of the field is cloud obscured.  The failed fields consist of 6 cloud affected images, the rest being observations of the galactic plane, causing field crowding.

\section{Data Calibration}
\subsection{Summary of pipeline}
The data pipeline used to analyse the resulting images is in the form of shell 
scripts which are used to call up the necessary software scripts.  The pipeline 
uses SExtractor \linebreak \citep{sextractor} to retrieve the data on the 
objects in each image, including their pixel position, flux, instrumental 
magnitude, background counts, isophotal area, FWHM, elongation, ellipticity, 
angle of the semi-major axis and the SExtractor flag data.  The setup of this 
has been fine-tuned to run optimally on both SkycamZ and SkycamT, with differing 
values for the pixel scale and gain parameters, to match those of each camera.

WCStools \citep{wcstools} is used to convert between pixel and world coordinates, using the WCS created by Astrometry.net, as SExtractor is incapable of processing the extra distortions added by the Astrometry.net software to account for the large field sizes.  

Table handling is performed using the Starlink Tables Infrastructure Library Tool Set (STILTS) \citep{STILTS}, \linebreak
which is a command line version of the popular Topcat program.  This provides the necessary controls for reading, modifying and linking tabular data produced in the pipeline. 

The main process behind the pipeline is the matching of objects detected in the fields by SExtractor to associated objects listed in the USNO-B catalogue \citep{USNOB}.  As the USNO-B catalogue provides all sky coverage down to a magnitude limit of 21 with photometric errors of $\pm$0.25, it provides photometric data for all the objects observable by the Skycams.  For use in the pipeline the USNO-B catalogue is cut to remove faint objects, which are below the detection threshold of the Skycams, this is to reduce the file sizes of the data in question but also to reduce the chances of erroneous object matching.  The limiting magnitude used for the cut is 18$^{th}$ and 12$^{th}$ magnitude for SkycamZ and SkycamT respectively.  Also for the catalogue data to be useful both B2 and R2 band data must be present to produce successful results.

\subsection{Calibration}
The first stage of analysing a night's data set is the calibration process, which is designed to align the photometry into a recognised filter set.  This is done for each individual night to accurately analyse the zeropoint for that particular night's observations.  The pipeline is  designed to calibrate the data using a set of 60 images, selected using a set of strict parameters to only analyse successfully taken images and not incomplete or distorted frames.  As the Skycams continuously take observations every minute, regardless of what the main telescope is doing, distorted images taken during slews and offsets are a common occurrence.  The frames are analysed consecutively from the first image taken during the night up until 60 images have been collected for further analysis.  For every image analysed, whether successful or not a line is added to a noticeboard file, which contains the results of the analysis for each image.  This is used to confirm which images have been utilised for producing the final calibrations and for what reasons the other images have not.  

The initial check performed is to confirm that the image has a WCS added to the image, as this is required to process the image further.  Also a check is performed to determine if the telescope has changed pointing since the last observation.  This is done to reduce the chance that a distorted field is passed through for analysis as if the telescope has moved since the previous observation there is no guarantee it has completely stabilised in its new position.  Therefore if the centre of the image has moved more than 1$^o$ since the previous image it is ignored.  Also if the previous image failed for not having a WCS, the current image is also ignored.  This is done to create a buffer around good, stable images whereby any images prior to the telescope stabilising or when the telescope begins to move to a new pointing, is ignored.  After these initial checks SExtractor is executed on the image and a list of sources is produced.

A count is performed on the SExtractor data to confirm how many objects are in the image.  Due to the Skycams operation schedule, images are taken at the opening of the dome.  This produces images of the dusk sky, which cannot be relied upon as being photometric.  Therefore any image with less than 200 or 600 objects, for SkycamZ and SkycamT respectively is flagged as a potential non-photometric image and is therefore ignored.  This method also removes frames which may suffer from cloud cover.  Further to this any object within 10 pixels of the field edge of the image is ignored to remove the chances of loss of signal off the end of the chip.

Once confirmed the image contains WCS data and an acceptable number of sources, checks are performed on the SExtractor data to determine the quality of the image. Very distorted fields cannot be assigned a WCS so are filtered out in the first stage, however as the astrometry fitting software is very robust, images which are distorted but the general positioning of objects are still determinable a WCS can be fitted.  To remove these images a set of checks are performed on the SExtractor output table data to determine if the image is distorted.  

Firstly, using the elongation data from the SExtractor output, which is the ratio of the semi-major and semi-minor axes lengths, a check is performed to determine if the image contains any object with an elongation ratio of greater than 12.  If this is the case the image is classed as distorted and no further processing is carried out.

Next a statistical calculation, known as excess kurtosis is calculated using the STILTS package \citep{STILTS}.  This provides an indication of how peaked a dataset is.  This is applied to the angle of the semi-major axis of the objects.  Using the results a large excess kurtosis value indicates heavily peaked data, which shows that the majority of objects are pointing in the same direction, which is a clear indicator of a distorted field.  For the Skycams this equates to any image with an object angle excess kurtosis $>$(-0.5) for SkycamZ and $>$(-0.45) for SkycamT.  Finally a combined check of elongation and object angle excess kurtosis is carried out, whereby if the excess kurtosis is $>$(-1.15) for SkycamZ or $>$(-0.8) for SkycamT and there are more than 2 objects with an elongation ratio of $>$5 for SkycamZ or 5 objects for SkycamT then the image is classed as distorted and it is ignored.  These criteria were defined empirically from study of the properties of actual Skycam images.

Due to the nature of the Skycams observing setup, with a set exposure time, on occasions the cameras will observe objects which will saturate the CCD.  This can affect the photometry of other objects.  Therefore with the target of a database of high quality data, any image containing a highly saturated object is ignored to save the integrity of the final results. This equates to any object with a total flux count of greater than 10$^7$ counts.

Also any object which has an associated SExtractor flag is ignored.  The flag data from SExtractor is used to warn of potential issues with the photometry, such as object blending and field edge objects.  Therefore to reduce the risk of contaminating the database with corrupt measurements, any object with a flag is ignored, however the image is not.

Upon completion of all the checks to remove distorted images WCStools is used to convert the pixel positions of the SExtractor data to celestial coordinates.  

With a complete set of WCS coordinates the objects can now be matched to corresponding USNO-B objects.  A data subset is produced from the USNO-B catalogue, which maps the area covered by the image in question, down to the upper magnitude limits of the instruments.  The two data sets are then crossmatched to pair up the observations to their equivalent USNO-B source data, using a tolerance of 9'' for SkycamZ and 148'' for SkycamT.  From the subsequent results the (B-R) colour is calculated for the objects from their USNO-B B2 and R2 magnitudes. Also calculated is the difference between the USNO-B R magnitude and the instrumental magnitude from the SExtractor data.

\begin{figure}[ht]
\centering
\includegraphics[width=0.34\textwidth, angle=270]{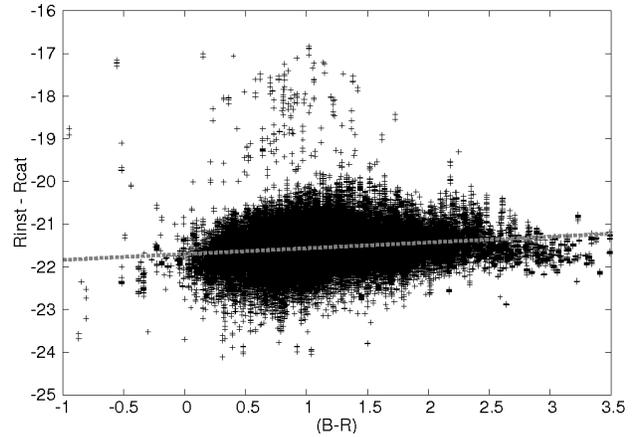}
\caption{An example calibration plot, with linear fit.  121,758 points are used to create the fit in this case.  This provides a fit with errors of 0.0116$\%$ and 1.58$\%$ on the intercept and gradient respectively.}
\end{figure}

Further quality control is performed to remove objects with a high or low colour index.  This is to remove objects where the data from the USNO-B catalogue is producing highly unusual colours, which are likely to be incorrect.  As these objects can have a significant affect on the final fit they are removed from the data set.  The colour limits used are -1$<$(B-R)$<$4.

Finally a second check for the number of observed objects is carried out, which only takes into account the number of matched objects. This will remove any images which have failed to match enough objects to reach the detection limit.  Upon completion of the analysis of a successful image the resultant data is added to a file containing the data from other successfully processed images from the same night.  When 60 images have been added the process stops and the results are used to produce a fit for the rest of the resulting night's data.

The results from the 60 images are used to plot the difference between the instrumental magnitude and the USNO-B R2 band magnitude against the (B-R) colour of the objects observed in the said images.  This is then analysed to produce a linear fit by the method of least square regression.  Figure 8 shows an example fit.  The results from the linear fit can then be used to calibrate data taken on the same night as the fit will account for both the standard offset of the system and the nightly variations in magnitude zeropoint, thus providing complete calibration data.

This system removes all but airmass errors from the data.  Due to the relatively large photometric errors caused by the uncertainties in catalogue magnitude and photometry, airmass will have a negligible effect on the data. This can be proven using the extinction data provided by the Isaac Newton Group (ING) group on La Palma's extinction with wavelength data \citep{extinction}.  This shows that at the peak throughput wavelength for the CCDs the extinction is 0.0991 mag per airmass, with the airmass at 35$^o$ being 1.740, which is usually the lowest pointing of the telescope. This gives a maximum airmass error of 0.07.  This is within the $\pm$0.1 uncertainty found through other causes of error in the system.

\subsection{Data analysis}
Upon completion of the calibration the main data analysis pipeline is initiated, which includes the reading and recording of the data from the images.  The data pipeline uses the same image quality checks as described in the calibration pipeline to remove distorted or potentially poor quality images.  Again WCStools is used to convert to celestial coordinates and USNO-B matching is performed.

Calibrated magnitudes (based on the USNO-B catalogue colour) and errors are calculated from the results of the linear fit, produced by the calibration pipeline for all matched objects.  Again colour cuts are used to filter out potentially incorrect catalogue data.  Also calculated are the sine and cosine of the declination for later use in the database.

After the matched data has been successfully handled the pipeline then moves on to deal with any detected sources that were not matched to USNO-B sources.  This can occur for many reasons, including; low quality astrometry data, lack of both colour magnitudes in the USNO-B catalogue, solar system objects or satellites and actual transient events.  Due to the possibility of these objects being real and either not previously catalogued in the USNO-B or simply not correctly matched to their subsequent catalogue data, the results from these objects are also recorded.

As these objects have not been matched to USNO-B sources they do not have a (B-R) colour, which is required for calibration.  Therefore an estimate of the median colour, (B-R)=1.5 is used.  Calibrated magnitudes and errors are then calculated for these sources using the same process.  As these objects have not been matched to USNO-B sources they are provided with a unique Skycam reference title based on their spatial coordinates.

Further checks are performed to remove any distorted images that may have passed through the initial round of image checks.  This is done by comparing the number of un-matched sources in the current image compared to the previous image, if the pointing remained the same.  If there are more than 1.5 times more unmatched sources then the image is corrupt in some way, most caused by the telescope slewing midway through an observation, producing a distorted field.  The data for this check is taken from the night's noticeboard file.

Finally the data from an image that passes all the quality checks is then stored in a custom MySQL database designed to house the data in its most optimal form.  The results stored in the database are the output from SeXtractor, with the spatial coordinates of the objects, the USNO-B data, which match to objects detected in the image and details from the FITS header of the image including the image coordinates, date and time of the image and its original file name.  

Upon the completion of processing a night's data, a \linebreak check is performed to determine the quality of the night's observations.  This is done by finding objects that have been observed $>$20 times that night.  The standard deviation of the calculated magnitudes of these objects is then calculated and the average standard deviation of all the objects that fit the observing parameters is calculated.  This provides a gauge of the quality of the night's observations.  A large standard deviation indicates a poor night's observations as sky variation will affect the results heavily relative to actual stellar variations in the observed objects.  This value is then added to the database for images taken during that night.  This can then be used at a later date to determine whether the results from that night can be considered photometric.

\subsection{Storage}
The resultant data from the analysis pipeline is stored in a MySQL database designed to maximise the potential of the system.  The structure of the database was designed for maximal access speed and minimal data redundancy. The data is stored across four tables, each aimed at a specific subset of the data.  The central table is designed to hold every data point of every object, this is crossmatched with an object based table which contains the name and coordinate details for each object in the database.  An image table contains the FITS header data from each image processed by the system. and finally a catalogue table contains the relevant details from the USNO-B catalogue for all matched objects.  Parts of the database are indexed to decrease the query time for commonly used queries.  The object reference title (USNOREF) is indexed in each table it appears, as this is widely used to find data on specific objects.  The Modified Julian date (MJD) is indexed in the image table, as this is regularly used to query by time with regards to images.  The nightly quality indicator (night$\_$std) is also indexed in the image table as nearly all queries with regards to object observations requires a cut on the quality of the data required.

\section{Data Performance}
\subsection{Efficiency}
For the system to run on the minutely timescale, the separate stages of data acquisition, reduction, analysis and storage must be capable of keeping up with this rate.  For a single exposure of 10 seconds it takes the system 50 seconds to complete the cycle of exposure, readout and all system overheads.

The instrumental reduction pipeline (bias, dark, \linebreak flat-fields etc.) runs on the archive computer, asynchronously from the data acquisition so as not to interrupt the imaging process.  A frame takes typically 4-40 seconds to process (median = 11 sec) with the variance entirely accounted for in the WCS fitting. As stated the image analysis pipeline does not currently run real time, however the final goal is to have the system running as the images are collated in the archive.  This requires that the pipeline is capable of running at a greater rate than 1 image per minute.  This is easily achieved with each image taking on average 10 seconds to complete.  This has also meant that to process a complete night's observations the pipeline only takes approximately 1.5 hours to complete on the archived data.

\begin{figure}[ht]
\centering
\includegraphics[width=0.4825\textwidth]{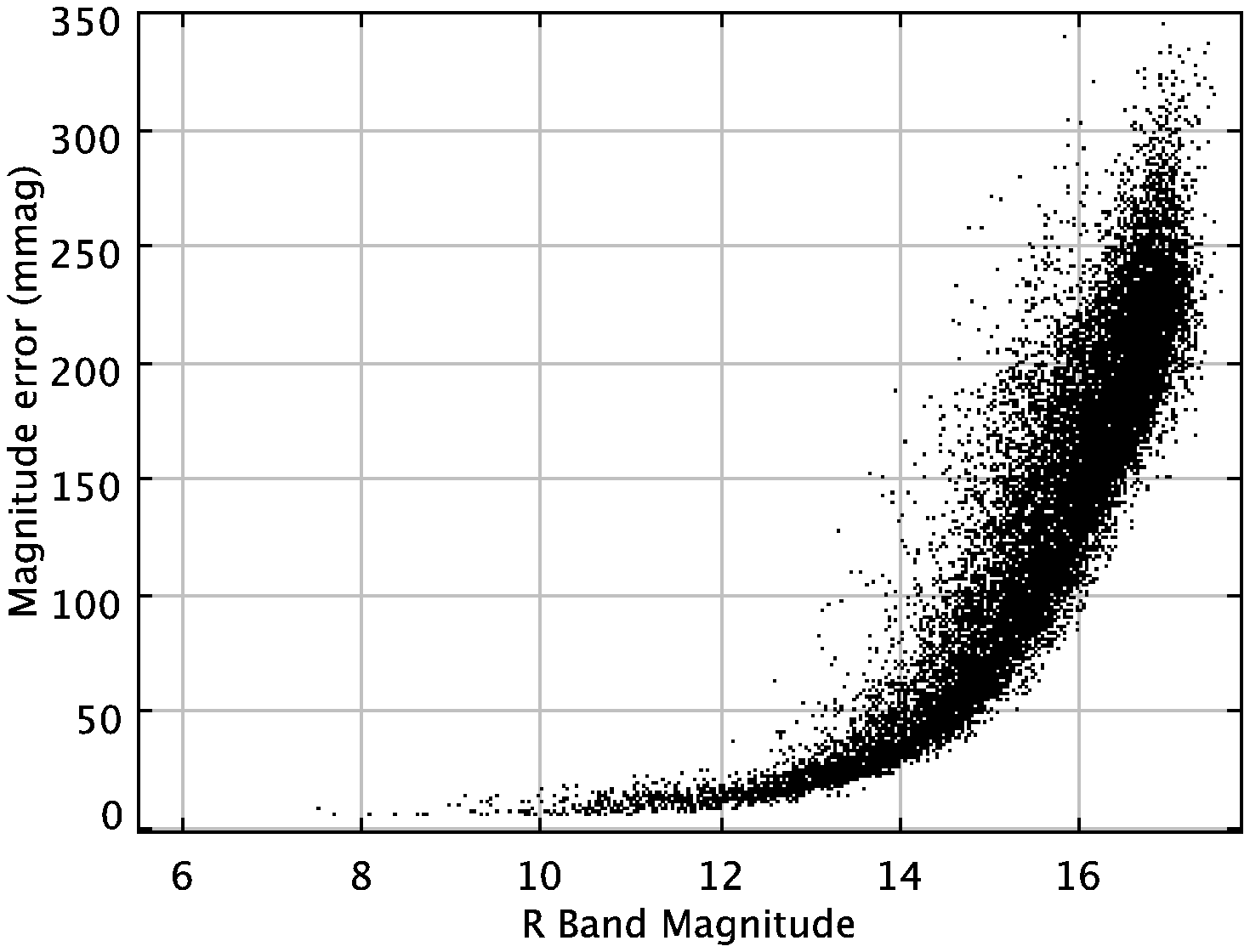}
\includegraphics[width=0.4825\textwidth]{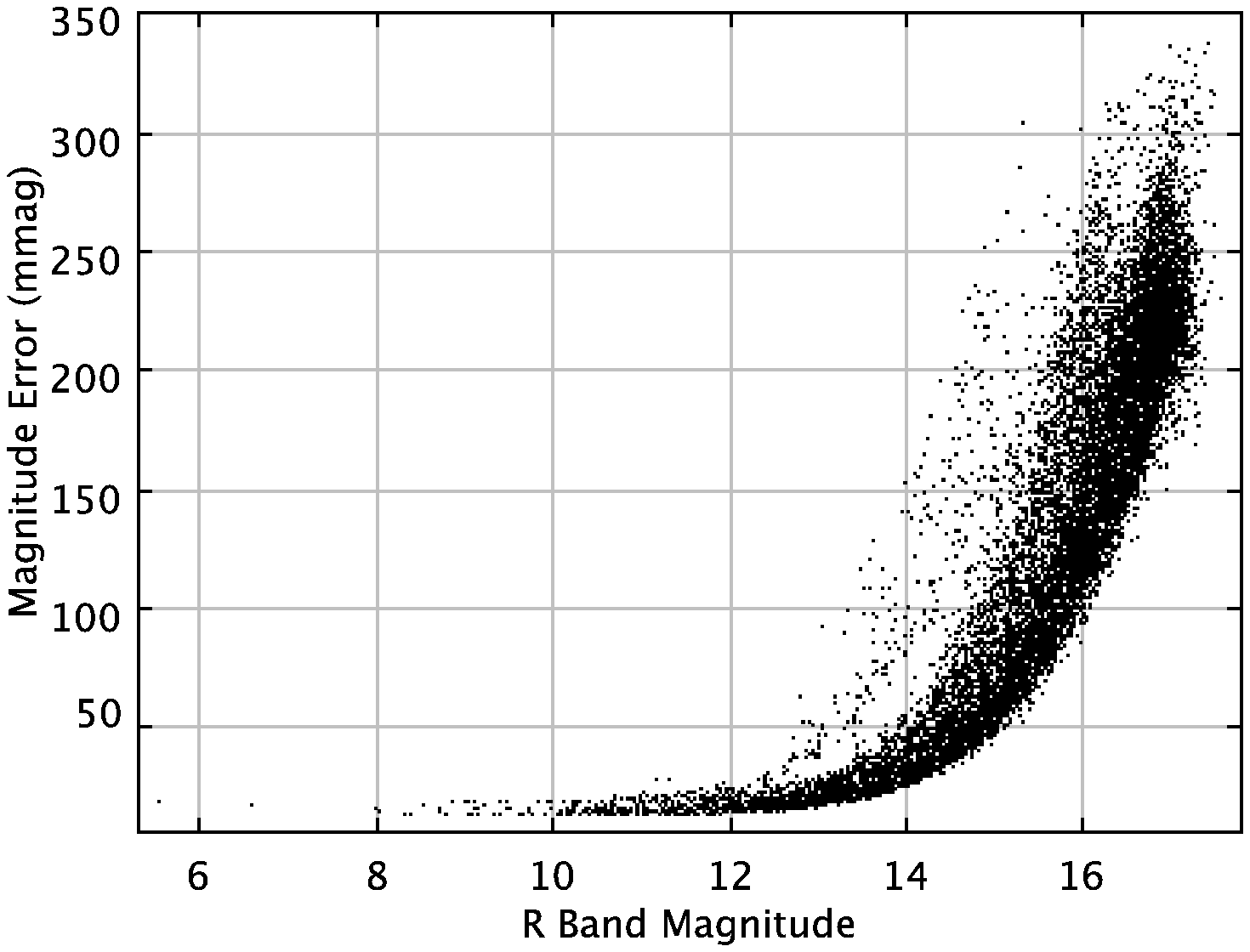}
\caption{Mean magnitude vs the error on the magnitude (in mmag) of sources in an example photometric night for SkycamZ, with (a) a single field shown (above) and (b) the night as a whole (below).}
\end{figure}

\begin{figure}[ht]
\centering
\includegraphics[width=0.4825\textwidth]{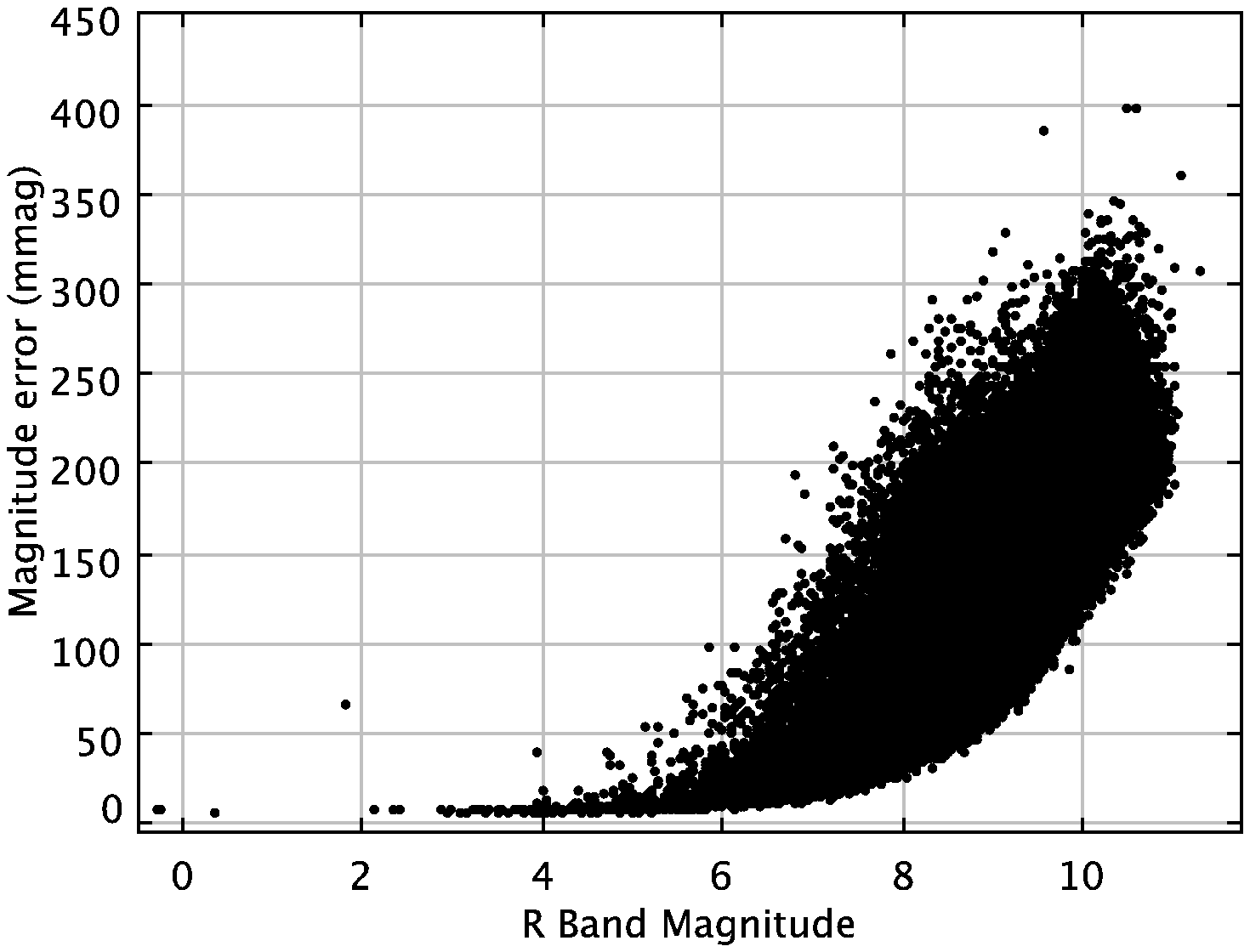}
\includegraphics[width=0.4825\textwidth]{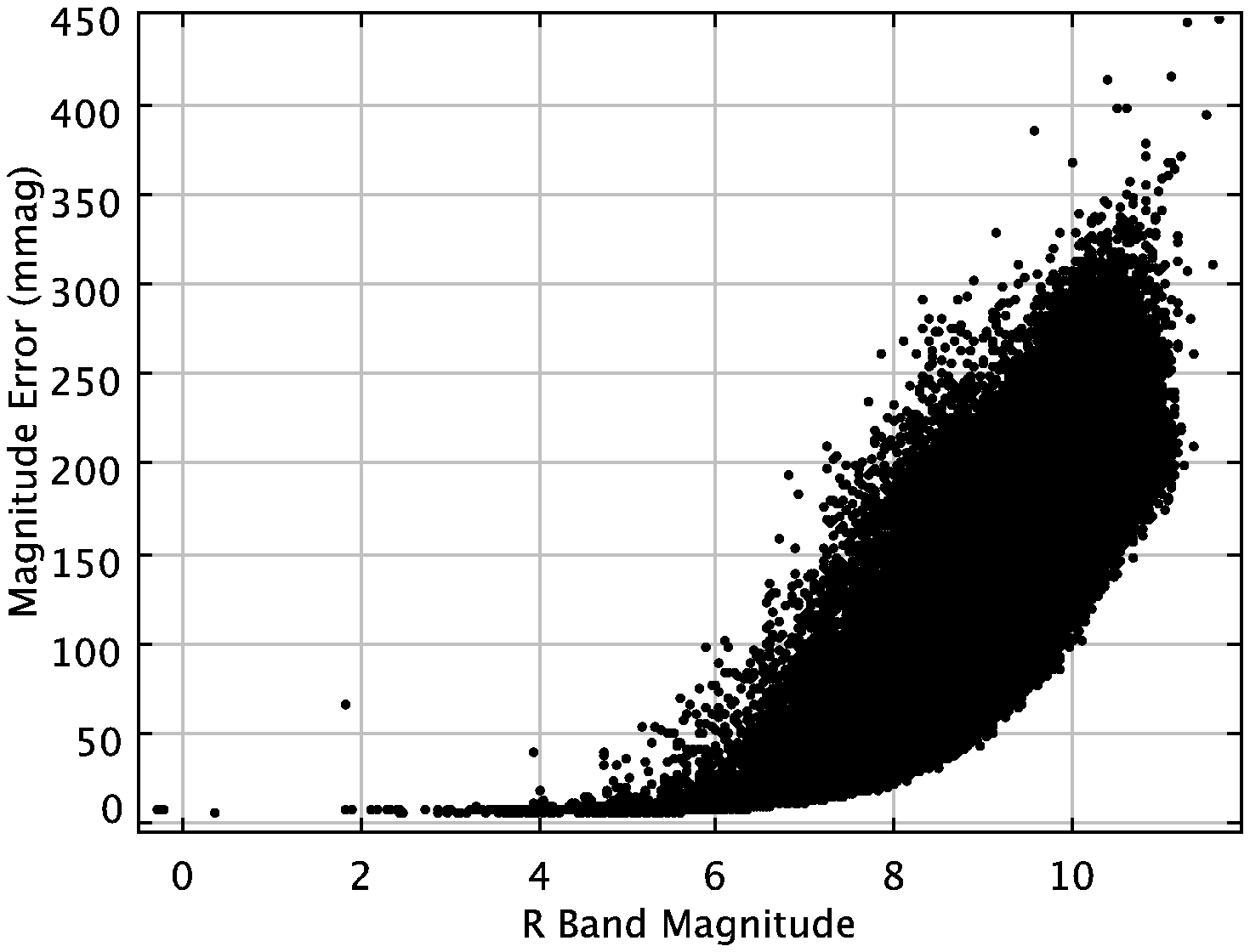}
\caption{Mean magnitude vs the error on the magnitude (in mmag) of sources in an example photometric night for SkycamT, with (a) a single field shown (above) and (b) the night as a whole (below).}
\end{figure}

\subsection{Photometric performance}
The photometric performance of the system can be split into two regimes; short and long term performance.  The performance over these regimes can be determined using photometric results held within the finalised database.

Short term stability of the instruments can be tested by looking at data from a single night.  By focussing on data of a single field, which has been observed many times in a night, it is possible to show how the photon counting errors vary with magnitude.  In this instance, other error sources, such as catalogue uncertainties and calibration errors will not be included.  Plotting the average magnitude and photometric error for each object observed within a single field, for a given night (as shown in Figure 9a and 10a) shows the relation of the photometric error with magnitude for each camera.  In comparison looking at the night as a whole (Figure 9b and 10b) introduces scatter from unknowns in the catalogue colours, as the colours are used in the calibration process (see Section 4.2).  By comparison the two data sets  are very similar in both size of the errors and their relation to the magnitude, therefore the scatter from the colours can be classified as a negligible effect.

Looking at the dataset as a whole introduces additional errors due to the calibration process as it is performed on a nightly basis.

Using magnitude binned data from the database provides an insight into how the errors in the system are formed (see Table 3 and 4).  The photon counting error from the object photometry, as used for short term stability investigations, is defined as the 'Magnitude Error'.  While the standard deviation of the magnitude of each single object, averaged for each magnitude bin is defined as the 'Magnitude S.D'.  This shows the overall error in the system, caused by both the photometric counting errors and errors brought in by the calibration process.  

By only focussing on the bright magnitude bins, where the photometric error is very small, the calibration error can be approximated as 0.16 in SkycamZ and 0.2 in SkycamT.  Using error propagation this can be confirmed for fainter magnitudes by combining the photometric and the calibration errors, which produces approximately the same results as the standard deviation values quoted.

The average standard deviation for the magnitude of all observed objects extracted from the database is 0.23 for SkycamZ and 0.27 for SkycamT.  This statistic will, invariably contain many, genuinely variable objects.

\begin{center}
\begin{table}[!ht]
\caption{SkycamZ: Binned magnitude data, containing the average photometric errors, average standard deviation of magnitude for each single object and the number of objects in each bin.}
\makebox[\linewidth]{
\begin{tabular}{|c|c|c|c|}
\hline
\multicolumn{3}{|c|}{Magnitude} & \multicolumn{1}{|c|}{Object} \\
Range & Error & S.D & count\\
\hline
4-5 & 0.0082 & 0.221 & 149\\ 
5-6 & 0.0081 & 0.174 & 388\\ 
6-7 & 0.0083 & 0.171 & 1091\\ 
7-8 & 0.0084 & 0.157 & 2984\\ 
8-9 & 0.0090 & 0.162 & 7520\\ 
9-10 & 0.0117 & 0.160 & 20113\\ 
10-11 & 0.0204 & 0.161 & 56480\\ 
11-12 & 0.032 & 0.159 & 142124\\ 
12-13 & 0.044 & 0.162 & 324485\\ 
13-14 & 0.070 & 0.173 & 701764\\ 
14-15 & 0.110 & 0.193 & 1273118\\ 
15-16 & 0.157 & 0.238 & 1744494\\ 
16-17 & 0.211 & 0.288 & 1430032\\ 
17-18 & 0.241 & 0.299 & 59647\\ 
\hline
\end{tabular}}
\end{table}
\end{center}

A final question is the presence of long term drifts in the calculated data.  Using all available photometric data for a set of standard stars, the long period stability of the system can be shown.  The results (Figure 11) show that there are no long period variations for the standard stars in either of the cameras, proving that the photometry from the cameras is stable.

\begin{center}
\begin{table}[!ht]
\caption{SkycamT: Binned magnitude data, containing the average photometric errors, average standard deviation of magnitude for each single object and the number of objects in each bin.}
\makebox[\linewidth]{
\begin{tabular}{|c|c|c|c|}
\hline
\multicolumn{3}{|c|}{Magnitude} & \multicolumn{1}{|c|}{Object} \\
Range & Error & S.D & count\\
\hline
-1-0 & 0.007 & 0.275 & 133\\
0-1 & 0.0093 & 0.206 & 254\\
1-2 & 0.0082 & 0.155 & 1629\\
2-3 & 0.0085 & 0.211 & 2685\\
3-4 & 0.0088 & 0.195 & 10746\\
4-5 & 0.011 & 0.200 & 38500\\
5-6 & 0.015 & 0.194 & 144335\\
6-7 & 0.025 & 0.208 & 490429\\
7-8 & 0.050 & 0.243 & 1427652\\
8-9 & 0.092 & 0.270 & 3881623\\
9-10 & 0.147 & 0.277 & 7027318\\
10-11 & 0.193 & 0.287 & 3933392\\
11-12 & 0.228 & 0.307 & 77657\\
\hline
\end{tabular}}
\end{table}
\end{center}

\subsection{Astrometric performance}
The astrometric performance of the system is difficult to quantify with the data produced from the pipeline.  Using the average separation between the sources extracted by SExtractor and the USNO-B catalogue object they have been linked to, it is possible to approximate the effect of false positive WCS fits on the data.  As the USNO-B catalogue has a very high pointing accuracy of 0.2'' at J2000 \citep{USNOB}, it is a useful yardstick by which the Skycam systems can be analysed.  The average separation between Skycam sources and USNO-B catalogue objects is 1.6939'' in SkycamZ and 28.888'' in SkycamT.  This shows that for both cameras the accuracy of the astrometric fit is better than half the pixel scale of the camera.

\begin{figure}[ht]
\centering
\includegraphics[width=0.25\textwidth, angle=270]{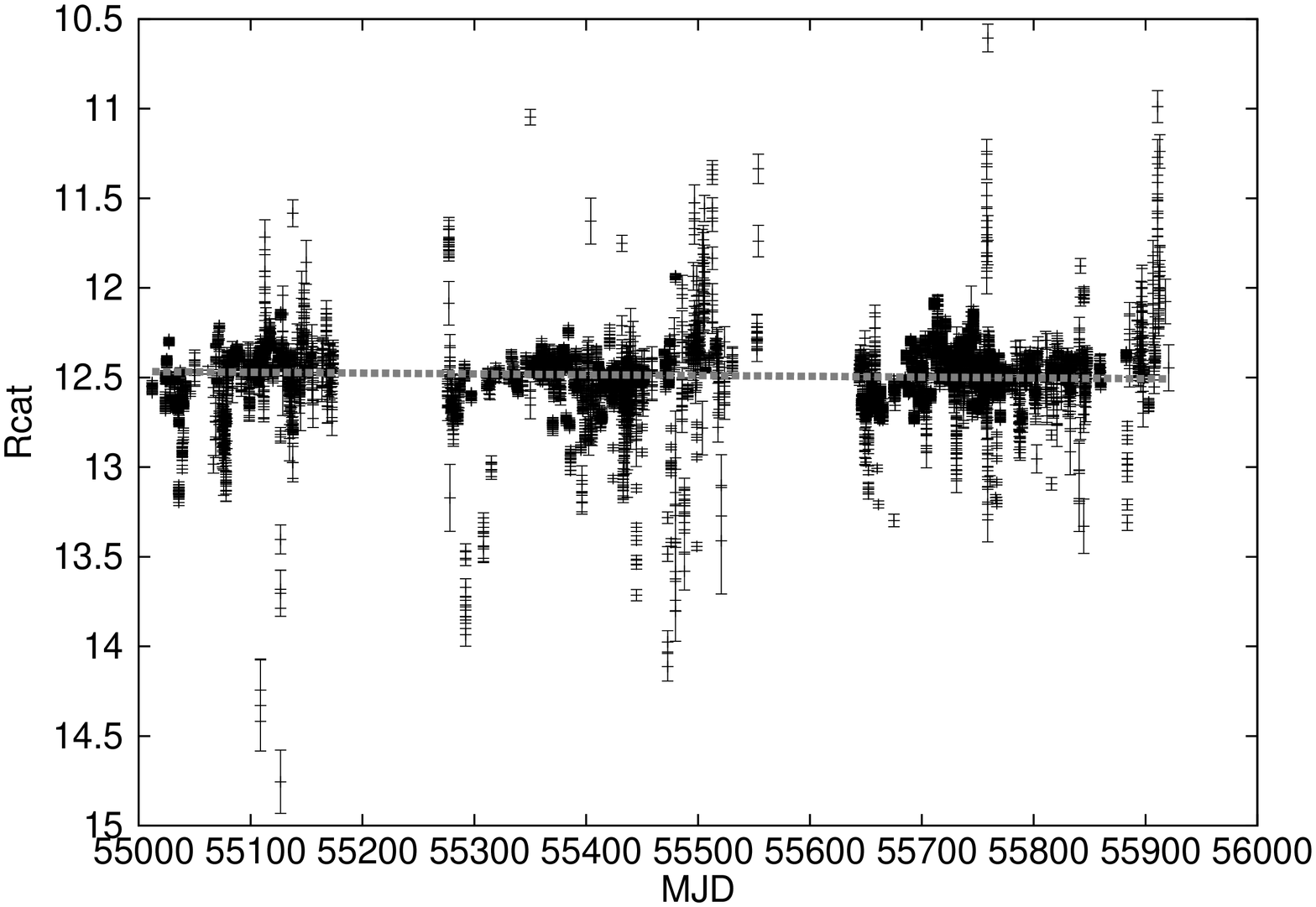}
\includegraphics[width=0.25\textwidth, angle=270]{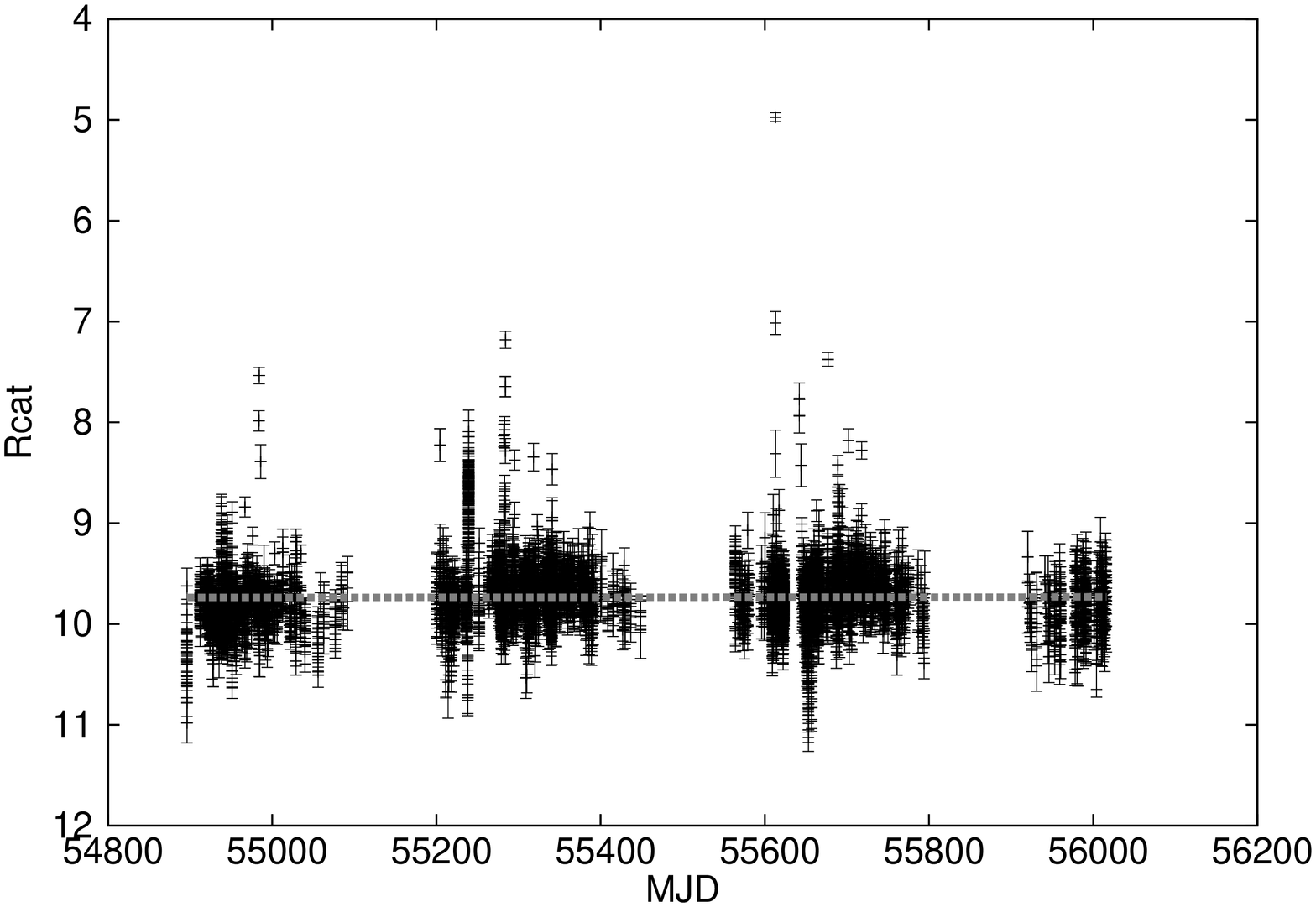}
\caption{Plots for standard star field objects covering the entire data collecting period. Plot 1 shows SkycamZ and plot 2 shows SkycamT. They display no long period trends.}
\end{figure}

\section{Summary}
We have commissioned and developed a set of autonomous wide fields camera systems attached to the Liverpool Telescope in La Palma.  The system has been in operation since March 2009 and the images up until March 2012 have been used to compile a database containing photometric data on all sources observed.  Consisting of a 1x1$^o$ and a 21x21$^o$ camera with photometric limits of 18$^{th}$ and 12$^{th}$ R band magnitude from observations of 10 second exposures at a rate of one per minute.  Observations continue to be made on a nightly basis and continue to be stored to the archive for later analysis.

We have shown the system to be photometrically stable, both in the short and long term using data from the resulting database and astrometrically stable, with average pointing accuracies of 1.7'' and 28.9'' compared to pixel scales of 3'' and 73''.

There is also the aim in the near future to activate the pipeline in a real-time basis so to provide up-to-the-minute data in the database and also the ability to trigger transient alerts, initially to the Liverpool Telescope and then the wider astronomy community.

There will be a further paper outlining the initial findings from the database, both of known catalogued sources and also previously un-registered variable objects along with the methods used to extract them from the database.

\textit{Acknowledgements. The Liverpool Telescope is operated on the island of La Palma by Liverpool John Moores University in the Spanish Observatorio del Roque de los Muchachos of the Instituto de Astrofisica de Canarias with financial support from the UK Science and Technology Facilities Council. This research made use of tools provided by Astrometry.net.  NRM acknowledges STFC for a postgraduate studentship during
which this work was undertaken.}

\bibliography{STILT_paper}

\end{document}